# Substance-like physical quantities in special relativity

Bernhard Rothenstein
"Politehnica" University of Timisoara, Physics Department
Timisoara, Romania

*Abstract. Handling substance-like physical quantities in the limits of special relativity theory we should make a net distinction between those which present a proper (rest) magnitude and those which have not. We show how the theory relates them via relativistic transformation equations.*

## 1. Introduction

Among the many physical quantities with which physicists operate we find certain class of physical quantities which play fundamental roles in classical and modern physics as well. They are called *substance-like* physical quantities and include: *mass, energy, momentum, electric charge* and *particle number density (amount of substance)*. Let *A* be the amount of such a substance-like physical quantity. They share in common the following properties:

-They are confined in bodies (in physical systems) and flow through space,
-Theirs distribution in a volume *V* is characterized by a *spatial density* $\rho_A$ given by

$$\rho_A = \frac{dA}{dV} \tag{1}$$

or

$$A = \int_V \rho_A dV \tag{2}$$

-Theirs flow through a given surface is characterized by the *current density* $\mathbf{j}_A$ given by

$$\mathbf{j}_A = \rho_A \mathbf{u}$$

**u** representing the velocity at which the flow takes place and by the *current* $I_A$ given by

$$I_A = \int_S \rho_A \mathbf{u} \cdot \mathbf{1_n} dS \tag{3}$$

*S* representing the surface through which the flow takes place, $\mathbf{1_n}$ the unit normal vector at a given point of the surface *S*

-They obey the *continuity equation*

$$\frac{dA}{dt} = I_A + \sum\nolimits_A \tag{4}$$

which shows that the speed of change of *A* located inside a given volume is due to a flow of A through its boundary surface ($I_A$) and to a production or destruction of *A* within the considered region of space ($\sum\nolimits_A$).

-For substance-like physical quantities that conserve $\sum\nolimits_A = 0$ and so

$$\frac{dA}{dt} = I_A \tag{5}$$

Theirs change in time being due to a flow of *A* through the boundary surface.[1]

## 2. Substance-like physical quantities in special relativity

Considering the problem in the limits of special relativity we state first that for mass (*m*), energy (*E*), particle number density *(n)* and spatial density ( $\rho$ ) we can define a proper magnitude ( $m_0, E_0, n_0, \rho_0$ ) measured by observers relative to whom the substance-like substance is confined is in a state of rest. Electric charge *q* has a special status because its magnitude is the same for all inertial observers in relative motion. For momentum, current density and current no proper values can be defined.

The involved inertial reference frames are K(XYZO) and K"(X'Y'Z'O'). The axes of the two frames are parallel to each other and the OX(O'X') axes are overlapped. K' moves with constant velocity *v* relative to K in the positive direction of the overlapped axes. At the common origin of time in the two frames, the origins O and O' are shortly located at the same point in space. In order to keep the problem as simple as possible, we consider that the body in which substance-like physical quantities are confined moves with constant velocity $u_x$ relative to K and with velocity $u_x'$ relative to K'. In accordance with the transformation of relativistic velocity we have

$$u_x = \frac{u_x' + V}{1 + \frac{Vu_x'}{c^2}}. \tag{6}$$

Relation (6) leads to the following obvious relativistic identities

$$\frac{1}{\sqrt{1 - \frac{u_x^2}{c^2}}} = \frac{1 + \frac{Vu_x'}{c^2}}{\sqrt{1 - \frac{V^2}{c^2}}\sqrt{1 - \frac{u_x'^2}{c^2}}} \tag{7}$$

$$\frac{u_x}{\sqrt{1 - \frac{u_x^2}{c^2}}} = u_x' \frac{1 + \frac{V}{u_x'}}{\sqrt{1 - \frac{V^2}{c^2}}\sqrt{1 - \frac{u_x'^2}{c^2}}}. \tag{8}$$

We also have the obvious identity

$$\frac{1 - \frac{u_x^2}{c^2}}{1 - \frac{u_x^2}{c^2}} = 1. \tag{9}$$

Multiply both sides of (8) and (9) with $m_0$. The result is

$$(\frac{m_0}{\sqrt{1 - \frac{u_x^2}{c^2}}}) = (\frac{m_0}{\sqrt{1 - \frac{u_x'^2}{c^2}}}) \frac{1 + \frac{Vu_x'}{c^2}}{\sqrt{1 - \frac{V^2}{c^2}}} \tag{10}$$

$$(\frac{m_0 u_x}{\sqrt{1 - \frac{u_x^2}{c^2}}}) = \frac{m_0 u_x'}{\sqrt{1 - \frac{u_x'^2}{c^2}}} \frac{1 + \frac{V}{u_x'}}{\sqrt{1 - \frac{V^2}{c^2}}}. \tag{11}$$

Physicists are experienced godfathers when they find out names for the physical quantities with which they operate. As long as they find out names for basic physical quantities used in order to construct a system of units (length, time, mass, Kelvin temperature and electric charge (no electric current) no special problems appear. Names given to physical quantities defined as a combination of some basic physical quantities generate fierce debates. Introducing the notations

$$m = \frac{m_0}{\sqrt{1 - \frac{u_x^2}{c^2}}} \qquad (12)$$

$$m' = \frac{m_0}{\sqrt{1 - \frac{u_x'^2}{c^2}}} \qquad (13)$$

physicists call *m* and *m'* *relativistic (dynamic) mass* measured by observers from K and K' respectively. Introducing the notations

$$p_x = \frac{m_0 u_x}{\sqrt{1 - \frac{u_x^2}{c^2}}} \qquad (14)$$

$$p_x' = \frac{m_0 u_x'}{\sqrt{1 - \frac{u_x'^2}{c^2}}}. \qquad (15)$$

Relativists call $p_x$ and $p_x'$ *relativistic momentum*. With those new notations (10) and (11) become

$$m = m' \frac{1 + \frac{V u_x'}{c^2}}{\sqrt{1 - \frac{V^2}{c^2}}} = \frac{m' + \frac{V}{c^2} p_x'}{\sqrt{1 - \frac{V^2}{c^2}}} \qquad (16)$$

$$p_x = p_x' \frac{1 + \frac{V}{u_x'}}{\sqrt{1 - \frac{V^2}{c^2}}} = \frac{p_x' + V m'}{\sqrt{1 - \frac{V^2}{c^2}}}. \qquad (17)$$

Multiplying both sides of (16) with $c^2$ (the invariance of *c* makes that it remains correct from the point of view of special relativity theory and introducing the notations

$$E = mc^2 = \frac{m_0 c^2}{\sqrt{1 - \frac{u_x^2}{c^2}}} \qquad (18)$$

$$E' = m'c^2 = \frac{m_0 c^2}{\sqrt{1 - \frac{u_x^2}{c^2}}} \qquad (19)$$

for the *relativistic energy* measured in K and in K' respectively it becomes

$$E = E' \frac{1 + \frac{Vu'_x}{c^2}}{\sqrt{1 - \frac{V^2}{c^2}}} = \frac{E' + Vp'_x}{\sqrt{1 - \frac{V^2}{c^2}}} \tag{20}$$

Whereas (17) can be presented as

$$p_x = \frac{p'_x + V \frac{E'}{c^2}}{\sqrt{1 - \frac{V^2}{c^2}}}. \tag{21}$$

If the body is at rest in K ($u_x = 0$) or at rest in K' ($u'_x = 0$) observers of the two frames measure its rest energy

$$E_0 = m_0 c^2. \tag{22}$$

As we see, special relativity theory relates two substance-like physical quantities for which we can define a proper value (mass) and for which such a concept is meaningless (momentum).

Considering a simple collision process between two particles from K and K' and working with the transformation equations for mass (energy) and momentum derived above we can show that relativistic mass (energy) and relativistic momentum lead to results in accordance with the conservation laws for mass (energy) and momentum as theirs classical counterparts do.[7]

Consider now *the spatial density* of *mass*. Its proper value is defined in its rest frame which

$$\rho_0 = \frac{m_0}{v_0} \tag{22}$$

where $m_o$ represents the proper mass uniformly distributed inside the proper volume $v_0$. As detected from K the spatial density of mass is

$$\rho = \frac{\rho_0}{\sqrt{1 - \frac{u_x^2}{c^2}}} \tag{23}$$

Whereas detected from K' it is

$$\rho' = \frac{\rho_0}{\sqrt{1 - \frac{u'^2_x}{c^2}}}. \tag{24}$$

The result is that the spatial density of mass transforms as

$$\rho = \rho' \left( \frac{1 - \frac{u'^2_x}{c^2}}{1 - \frac{u_x^2}{c^2}} \right) = \left( \frac{1 + \frac{Vu'_x}{c^2}}{\sqrt{1 - \frac{V^2}{c^2}}} \right) \frac{\rho' + \frac{v}{c^2} j'_x}{\sqrt{1 - \frac{V^2}{c^2}}}. \tag{25}$$

Multiplying both sides of (25) with $u_x$ we obtain that the current density transforms as

$$j_x = \rho u_x = \rho' \frac{(u'_x+V)(1+\frac{Vu'_x}{c^2})}{1-\frac{V^2}{c^2}} = \frac{1+\frac{Vu'_x}{c^2}}{\sqrt{1-\frac{V^2}{c^2}}} \frac{j'_x+V\rho'}{\sqrt{1-\frac{V^2}{c^2}}} = \frac{u'_x+V}{\sqrt{1-\frac{V^2}{c^2}}} \frac{\rho'+\frac{V}{c^2}j'_x}{\sqrt{1-\frac{V^2}{c^2}}} \quad (26)$$

Relations (25) and (26) show how special relativity relates a physical quantity with proper value ($\rho$) to a physical quantity that has no proper value ($j_x$).

The situation becomes simpler in the case of the substance like physical quantity electric charge $q$ that as we have mentioned above is a relativistic invariant. Its spatial density is

$$\rho = \frac{q}{v_0\sqrt{1-\frac{u_x^2}{c^2}}} \quad (27)$$

in K and

$$\rho' = \frac{q}{v\sqrt{1-\frac{u'_x{}^2}{c^2}}} \quad (28)$$

in K' resulting that it transforms as

$$\rho = \rho' \frac{1+\frac{Vu'_x}{c^2}}{\sqrt{1-\frac{V^2}{c^2}}} = \frac{\rho'+\frac{V}{c^2}j'_x}{\sqrt{1-\frac{V^2}{c^2}}}. \quad (29)$$

We define the proper particle number density (amount of substance) as

$$n_0 = \frac{N}{v_0} \quad (30)$$

where the counted number of stable particles N is a relativistic invariant. The experience gained so far tell us that $n$ and $n'$ measured in K and in K' respectively transform as

$$n = n' \frac{1+\frac{Vu'_x}{c^2}}{\sqrt{1-\frac{V^2}{c^2}}}. \quad (31)$$

Using (31) we can recover some of the transformation equations derived above.

    It is interesting to mention that we can define substance-like physical quantities in empty space as well. Consider that there is a region in empty space where we have an uniform electric field intensity showing in the positive direction of the $O^0Y^0$ axis

$$E_y^0 = \frac{Q}{2\varepsilon_0 A^0} \quad (32)$$

Generated by an uniformly distributed electric charge on a very large surface $A_0$ confined in the $X^0O^0Z^0$ plane of its rest frame $K^0(X^0Y^0Z^0O^0)$ that moves with velocity $u_x$ relative to K and with velocity $u_x'$ relative to K'. The same electric field intensity detected from K is

$$E_y = \frac{Q}{2\varepsilon_0 A^0 \sqrt{1 - \frac{u_x^2}{c^2}}} \qquad (33)$$

whereas detected from K' it is

$$E_y' = \frac{Q}{2\varepsilon_0 A^0 \sqrt{1 - \frac{u_x'^2}{c^2}}} \qquad (34)$$

leading to the transformation equation

$$E_y = E_y' \frac{1 + \frac{Vu_x'}{c^2}}{\sqrt{1 - \frac{V^2}{c^2}}} = \frac{E_y' + \frac{Vu_x'}{c^2} E_y'}{\sqrt{1 - \frac{V^2}{c^2}}}. \qquad (35)$$

At that point of the derivations the physicist introduces the notation

$$\mathbf{B}_z' = \frac{\mathbf{u}_x' \times \mathbf{E}_y'}{c^2} \qquad (37)$$

with which (35) becomes

$$E_y = \frac{E_y' + VB_z'}{\sqrt{1 - \frac{V^2}{c^2}}} \qquad (39)$$

$B_z$ ($B_z'$) transforming as

$$B_z = \frac{u_x E_y}{c^2} = E_y' \frac{u_x' + V}{c^2 \sqrt{1 - \frac{V^2}{c^2}}} = \frac{B_z' + \frac{V}{c^2} E_y'}{\sqrt{1 - \frac{V^2}{c^2}}}. \qquad (40)$$

The density of the uniformly distributed energy is

$$\rho = \varepsilon_0 E_y^2 \qquad (41)$$

in K and

$$\rho' = \varepsilon_0 E_y'^2 \qquad (42)$$

in K' that transforms as described by (25). The energy carried through the unit normal surface in unit time transforms as described by (26).

In a plane electromagnetic wave propagating through empty space ($u_x = u_x' = c$) the relations derived above become

$$E_y = kE_y' \qquad (43)$$

$$B_z = kB_z'  \qquad (44)$$
$$\rho = k^2 \rho' \qquad (45)$$
$$j_x = k^2 j_x' \qquad (46)$$

where

$$k = \sqrt{\frac{1 + \dfrac{V}{c}}{1 - \dfrac{V}{c}}}. \qquad (47)$$

We return now to the identity (10). Multiplying both its sides with the proper value of a substance-like physical quantity say with $E_0^2$ it leads to

$$E^2 - c^2 p_x^2 = E_0^2 . \qquad (48)$$

Multiplying both sides of (10) with $\rho_0^2$ it leads to

$$c^2 \rho^2 - j_x^2 = c^2 \rho_0^2 . \qquad (49)$$

Relations (48) and (49) show the way in which substance-like physical quantities with and without proper values lead to invariant expressions that hold in all inertial reference frames.

### 3. Conclusions

We have shown that handling substance-like physical quantities in the limits of special relativity theory we should make a net distinction between those that have a proper amount and those that have not. Special relativity reveals the way in which transformation equations relate mass (energy) and momentum and volume density to current density.

### Reference

[1]G. Bruno Schmid, "An up-to-date approach to physics," Am. J .Phys. **52,** 794-799 (1984) and reference herein.